\documentstyle[12pt]{article}
\input{psfig}

\newcommand{\beq}{\begin{equation}}
\newcommand{\eeq}{\end{equation}}
\newcommand{\bea}{\begin{eqnarray}}
\newcommand{\eea}{\end{eqnarray}}

\begin{document}
\title{{\bf Strange~/~anti-strange asymmetry in the nucleon sea} }
\author{{H.~R. Christiansen}\thanks{e-mail: hugo@cat.cbpf.br}
\ {\small and} { J. Magnin}\thanks{e-mail: jmagnin@lafex.cbpf.br} \\  
{\normalsize Centro Brasileiro de Pesquisas F\'{\i}sicas}, 
{\small CBPF - DCP} \\ 
{\normalsize Rua Dr. Xavier Sigaud 150, 22290-180, 
Rio de Janeiro, Brazil}}

\date{{\tt To appear in Phys.~Lett.~B}}
\maketitle

\begin{abstract}
We analyze the non-perturbative 
structure of the strange sea of the nucleon within a 
meson cloud picture. 
In a low $Q^2$ approach in which the nucleon is viewed as a 
three valon bound state, we evaluate the 
probability distribution of an in-nucleon Kaon-Hyperon pair 
in terms of splitting functions and recombination. 
The resulting kaon and hyperon probability densities are convoluted with 
suitable strange distributions inside the meson and baryon in order
to obtain non-perturbative contributions to the strange sea of the 
nucleon. We find a structured strange/anti-strange asymmetry,
displaying a clear excess of quarks (anti-quarks) for large (small)
momentum fractions. 
\end{abstract}

%%%%%%%%%%%%%%%%%%%%%%%%%%%%%%%%%%%%%%%%%%%%%%%%%%%%%%%%%%%%%%%%%%%%%%
\section*{Motivation}

According to the Quark Parton Model, with its subsequent 
improvements coming from Quantum Chromodynamics, it is known that 
hadrons are built up from a fixed number of valence quarks plus 
a fluctuating number of gluons and sea quark anti-quark pairs.
As the momentum scale rises up, the nucleon's sea is mainly perturbatively 
generated and consequently quarks and anti-quarks have the same 
probability and momentum distributions. However, a small fraction 
of the sea may be associated with non-perturbative processes raising 
the possibility of generating {\it unequal} quark and anti-quark 
distributions.
For instance, a non-perturbative $q$-$\bar{q}$ asymmetry in the  
charmed sea of the nucleon could play an important role in explaining 
the excess of events at large $x$ and $Q^2$ in $e^+p$ neutral 
and charged current deep inelastic scattering \cite{mel-tho}, 
as seen at HERA by H1 \cite{h1} and ZEUS \cite{zeus}.  
Regarding the strange nucleon's sea, the present 
status of the experimental data does not exclude the possibility of 
asymmetric $s$ and $\bar{s}$ distributions. Actually, it is an 
important question related to current experimental
research. 
Recently,  the CCFR Collaboration has analyzed the 
strange quark distribution in nucleons 
allowing explicitly for $s(x) \ne \bar{s}(x)$ distributions 
\cite{ccfr} (see also the CTEQ analyses \cite{cteq}). 
Although the analysis of CCFR seems to indicate a rather small
asymmetry, the error bars are still too large to be conclusive, leaving
enough place to conjecture about sizeable unequal $s$ and $\bar{s}$ 
distributions (see discussion in Refs.~\cite{signal,bro-ma}).  
Most recently, the E866 Collaboration \cite{e866} 
has measured a large $\bar{u}$-$\bar{d}$ asymmetry in the nucleon 
which may be succesfully accomodated in the framework of a  
pion cloud model.
 
The question of a non-perturbative asymmetry in the nucleon sea has
been addressed by a number of authors.
The most significant framework to discuss this issue appears to be
the Meson Cloud Model (MCM). In this context,
it is assumed that the nucleon can fluctuate to a meson-baryon
bound state with a small probability. Non-perturbative 
sea-quark distributions are then associated with 
the valence quark and anti-quark distributions
inside the baryon and the meson respectively. A first attempt
in this direction was carried out in \cite{signal} where the 
nucleon-meson-baryon vertex is parameterized with ad-hoc form factors.
In Ref.~\cite{bro-ma}, intrinsic sea quark and anti-quarks are rearranged 
with valence quarks using simple two-body wave functions for describing
meson-baryon bound states. 
Finally, let us also mention a different approach to the 
problem, based on a generalized chiral Gross-Neveu model at large 
$N_c$ \cite{burkardt}. 
%Finally, let us also mention the approach of 
%Ref.~\cite{chinos}, where  the different interactions
%experienced by the perturbatively generated quark and anti-quark 
%are simulated by giving different masses to each one, consequently 
%producing different distributions for sea quark and anti-quarks.

In this letter, we employ both effective and perturbative degrees of 
freedom and show that a well-known
scheme related to hadron production in hadronic collisions, viz. 
recombination, may be succesfully applied for describing meson-baryon 
fluctuations yielding non-perturbative sea quark-antiquark asymmetries. 
Our description of the nucleon-meson-baryon vertex starts
from the perturbative production of a $q\bar{q}$ pair out of a
dressed valence quark. Such perturbative sea quarks subsequently 
rearrange with the remaining valence quarks {\it via} recombination,
in order to give rise to a hadronic nucleon fluctuation. 
As we will see, this provides a phenomenologically
motivated one-parameter dependent description of the low $Q^2$ scale
meson and baryon probability distributions inside the nucleon. 
The non-perturbative contributions to the strange sea of the nucleon 
are then calculated by means of widely used convolution forms. We obtain
a structured asymmetry indicating a clear excess of strange quarks 
(antiquarks) for large (small) momentum fraction, becoming negligible 
beyond $x \sim 0.8$.

%%%%%%%%%%%%%%%%%%%%%%%%%%%%%%%%%%%%%%%%%%%%%%%%%%%%%%%%%%%%%%%%%%%%%%%
 
\section*{Strange meson-baryon fluctuation and $s-\bar{s}$ asymmetry 
in the nucleon}

We start by considering a simple picture of the nucleon in the 
infinite momentum frame as being formed by three dressed valence 
quarks - {\it valons} - which carry all of its momentum. 
The valon distribution in the nucleon has been calculated by 
Hwa in Ref.\cite{hwa}, and it has been found to be
\begin{equation}
v(x) = \frac{105}{16} \sqrt{x} \left( 1 - x\right)^2.
\label{eq1}
\end{equation}
For the sake of simplicity, we will not distinguish between
$u$ and $d$ valon densities, so identical momentum distribution will be
assumed for both flavors.

In the framework of the MCM, the nucleon can fluctuate to a  
meson-hyperon bound state carrying zero net strangeness. As a first step
in such a process, we may consider that  
each valon can emit a gluon which, before interacting, 
decays perturbatively into a $s\bar s$ pair. The probability of 
having such a perturbative $q\bar{q}$ pair can then be computed in terms 
of Altarelli-Parisi splitting functions \cite{alta-par} 
\beq
 P_{gq} (z) = \frac{4}{3} \frac{1+(1-z)^2}{z},\ \ \ \ 
P_{qg} (z) = \frac{1}{2} \left( z^2 + (1-z)^2 \right).
\label{eq2}
\eeq
These functions have a physical interpretation as the probability 
of gluon emision and $q\bar{q}$ creation with momentum fraction $z$ 
from a parent quark or gluon respectively. Hence, 
\begin{equation}
q(x,Q^2) = \bar{q}(x,Q^2) = N \frac{\alpha_{st}^2(Q^2)}{(2\pi)^2}
\int_x^1 {\frac{dy}{y} P_{qg}\left(\frac{x}{y}\right) 
\int_y^1{\frac{dz}{z} P_{gq}\left(\frac{y}{z}\right) v(z)}}
\label{eq3}
\end{equation}
is the joint probability density of obtaining a quark or anti-quark 
coming from  subsequent decays $v \rightarrow v + g$ 
and $g \rightarrow q + \bar{q}$ at some fixed low $Q^2$. 
As the valon distribution does not depend on $Q^2$ \cite{hwa}, 
the scale dependence in eq.~(\ref{eq3}) only exhibits through the 
strong coupling constant $\alpha_{st}$. The range of values of $Q^2$ 
at which the process of virtual pair creation occurs in our approach
is typically about 1 GeV$^2$, as dictated by 
the valon model of the nucleon. For definiteness, 
we will use $Q = 0.8$ GeV as in Ref.~\cite{hwa}, for which 
$\alpha_{st}^2 \sim 0.3$ is still sufficiently small to allow for 
a perturbative evaluation of the $q\bar{q}$ pair production.
Since the scale must be consistent with the valon picture,
the value of $Q^2$ is not really free and cannot be used
to control the flavor produced at the $gq\bar{q}$ 
vertex. Instead, this role can be ascribed to the normalization constant
$N$, which must be such that to a heavier quark corresponds a lower 
value of $N$.
For instance, it could be fixed by comparing the output of eq.~(\ref{eq3}) 
with fit analyses of experimental data of sea quark distributions
in the nucleon at low $Q^2$. However, we will for simplicity include 
the whole factor appearing in eq.~(\ref{eq3}) in the global normalization 
of the $\left|MB\right>$ Fock state.

Once a $s\bar{s}$ pair is produced, it can rearrange itself with the 
remaining valons so as to form a most energetically favored 
meson-baryon bound state. To obtain the meson and baryon probability 
densities inside the nucleon, one has to employ effective techniques 
in order to deal with the non-perturbative {\em QCD} processes inherent to 
the dressing of quarks into hadrons. Although in-nucleon meson and 
baryon are virtual states, 
one may assume that the mechanisms involved in their formation are similar 
to those at work in the production of real hadrons in hadronic collisions. 
Therefore we shall proceed to describe the $N\rightarrow MB$ fluctuation by
means of a well-known recombination model approach \cite{das-hwa}.

Notice that as the nucleon fluctuates 
into a meson-baryon bound state, the meson and baryon distributions 
inside the nucleon are not independent. Actually, to ensure the zero 
net strangeness of the nucleon and momentum conservation, the in-nucleon
meson and baryon distributions must fulfill two basic constraints:
\beq
\int_0^1 {dx \left[P_B(x) - P_M(x) \right]} =  0, \ \ \
\int_0^1 {dx \left[xP_B(x) + xP_M(x) \right]}  =  1,
\label{condiciones}
\eeq
which can be both guaranteed by choosing
\begin{equation}
P_M(x) = P_B(1-x)
\label{eq4}
\end{equation}
for all momentum fractions $x$.\footnote{Although
condition (\ref{eq4}) has been generally employed
(see {\it e.g.} Refs.~\cite{mel-tho,malheiro}), it is not a unique choice
since eqs.~(\ref{condiciones}) relate the integrals of the meson and baryon 
probability densities and not the distributions themselves.}
In addition to satisfying conservation laws, 
condition (\ref{eq4}) enables us to calculate just one of the two 
distributions. Therefore, we can proceed to compute 
the strange meson distribution $P_M(x)$  
along the lines of Ref.~\cite{das-hwa} and then relate it to the hyperon
probability as indicated in eq.~(\ref{eq4}). 
Using the recombination model, the probability density 
of having a  meson out of two quarks is given by
\begin{equation}
P_M (x) = \int_0^x \frac{dy}{y} {\int_0^{x-y}\frac{dz}{z} F(y,z) R(x,y,z)},
\label{eq5}
\end{equation}
where $F(y,z)$ is 
\begin{equation}
F(y,z) = \beta\, yv(y)\, z\bar{q}(z) (1-y-z)
\label{eq6}
\end{equation}
and $R(x,y,z)$ is the recombination function associated
with the meson formation 
\begin{equation}
R(y,z) = \alpha\frac{yz}{x^2} 
\delta \left(1 - \frac{y+z}{x}\right) \; .
\label{eq7}
\end{equation}
The normalization of the meson distribution
is not given by the recombination model. We shall thus
fix the overall normalization as given by the experimental
probability that the 
strange hadronic fluctuation occurs; this value is phenomenologically
estimated to be about $4-10 \%$ (see {\it e.g.} 
Refs.\cite{bro-ma, malheiro}).
In Fig.1 we show our calculated meson and 
baryon probability densities inside the nucleon. 

The non-perturbative strange and anti-strange sea distributions 
can be now computed by means of the two-level convolution formulas  
\beq
s^{NP}(x) = \int^1_x {\frac{dy}{y} P_B(y)\ s_{B}(x/y)}\ \
\ \ \ \bar{s}^{NP}(x) = \int^1_x {\frac{dy}{y} P_M(y)\ \bar{s}_{M}(x/y)},
\label{eq8}
\eeq
where the sources $s_{B}(x)$ and $\bar{s}_{M}(x)$ are primarily the 
probability densities of the strange valence quark and anti-quark 
in  baryon and meson respectively, evaluated at the hadronic 
scale $Q^2$ \cite{signal}.
In principle, to obtain the non-perturbative distributions given by 
eqs.~(\ref{eq8}), one should sum over all the meson-hyperon 
fluctuations of the nucleon but, since such hadronic Fock states are 
necessarilly off-shell, the most likely configurations are 
those closest to the nucleon energy-shell, namely 
$\Lambda^0K^+$, $\Sigma^+K^0$ and $\Sigma^0K^+$, for a 
proton state.
However, as we are using the same distribution for $u$ and $d$ valons, 
the three configurations above contribute in the same way to the 
strange non-perturbative structure of the nucleon. Other fluctuations 
involving heavier in-nucleon mesons and baryons should be strongly 
suppressed due to their high virtuality \footnote{For instance, a 
loosely bound $p\phi$ configuration 
can be neglected since it is rather heavier than a 
$\Lambda K$ state and Zweig's rule suppressed.}.

As long as experimental measurements are lacking at present,
several choices are possible for the  $\bar{s}_M$ and $s_B$ 
distributions in $K$-ons and $\Lambda$ or $\Sigma$ baryons respectively. 
In this respect, it has been a common practice to employ modified light 
valence quark distributions of pions and protons \cite{signal,malheiro}. 
However, we think that one should better use simple forms reflecting
the fact that strange valence quarks should carry a rather large 
amount of momentum in $S=\pm 1$ hadron states at low $Q^2$ scales. 
Indeed,  the momentum distribution found by Shigetani 
{\it et al.} \cite{shigetani} in the framework of a Nambu-Jona 
Lasinio model at low $Q^2$, 
adequately reflects this feature (see also Ref.\cite{kaon}). 
A similar 
form for the strange distribution in $K$-ons has been
found very recently in \cite{suecos}, 
using a Monte Carlo based program for generating valence distributions in 
hadrons. We thus put forward the following simple form 
\beq
x\bar{s}_M(x)=6 x^2 (1-x)
\label{eq10}
\eeq
which is an excellent approximation to these phenomenological data.
For similar reasons, in $\Lambda$ and $\Sigma$ baryons we 
expect a $s_B$ momentum distribution peaked around $1/2$, which 
after normalization can be approximated by
\beq
xs_B(x)=12 x^2 (1-x)^2 \; .
\label{eq11}
\eeq

For comparison, we have also considered other forms for $\bar{s}_M$ 
and $s_B$ 
in order to display their effect on the shape of 
the non-perturbative distributions of eqs.~(\ref{eq8}). As suggested
in \cite{signal,malheiro}, we have used $\bar{s}_M = (1-x)^{0.18} q^{\pi}$, 
where $q^{\pi}$ is the valence distribution in pions, and 
$s_B = u_N/2$, with $u_N$ the up valence distribution in the nucleon; 
$q^\pi$ and $u_N$ are those given in Ref.\cite{suecos}.
In Fig.~2 we show the $s^{NP}$ and $\bar{s}^{NP}$ distributions 
as well as the $s-\bar{s}$ asymmetry predicted within our model. 

%%%%%%%%%%%%%%%%

\section*{Summary and discussion}

By means of an approach involving both effective and perturbative 
degrees of freedom, we have obtained $s$ and $\bar{s}$ non-perturbative 
distributions in the nucleon's sea. 
 
We based our approach on a valon description of the initial state 
of the nucleon, which perturbatively produces sea quark/anti-quark pairs. 
Thereafter these quarks and anti-quarks give rise to a hadronic bound 
state by means of recombination with the remaining valons. 
An interesting feature of the model is that it allows a full 
representation of the
non-pertur\-ba\-tive processes inside the nucleon in terms of a well-known 
effective scheme of hadron physics \cite{das-hwa}. 
In this way, a specific connection between the physics of hadronic reactions 
and that of hadron fluctuations is established in the approach. 
This is a pleasant aspect 
of the approach since the same physical principles, related to 
hadronization, should be involved in both situations.
The present model also exhibits a remarkable economy of parameters when 
compared to other approaches in the literature. Notice for instance that 
neither  nucleon-meson-hyperon coupling constants nor vertex functions  
are used to recognize the extended nature of the nucleon-meson-hyperon 
vertex, avoiding the need of such parameters, whose values
are still controversial.

As can be seen in Fig.~2, the model predicts a definite 
structured asymmetry in the strange sea of the nucleon. 
As we have discussed in the 
introduction, our results are compatible with the present status of 
experimental data coming from $\nu N$ and $\bar\nu N$ scattering,
due to the large error bars allowed for such 
distributions \cite{ccfr}. 
One should however notice that the predicted non-perturbative 
$s^{NP}$ and $\bar{s}^{NP}$ distributions depend on the form of the 
strange and anti-strange distributions in the in-nucleon 
baryon and meson respectively. So, although the model is reliable in 
predicting different distributions for sea quarks and anti-quarks, their 
exact shapes cannot be determined until we have more confident results for  
quark densities inside strange hadrons. Nevertheless, the model 
clearly predicts an excess of sea quarks over anti-quarks  
carrying a large fraction of the nucleon's momentum (see Fig.~2) and
this appears to be independent of the exact form of the strange 
and anti-strange distributions inside the in-nucleon baryon and meson. 
This result coincides with the predictions of 
Ref.~\cite{signal,bro-ma,burkardt} 
and strongly differs with the outcome of Ref.~\cite{malheiro}. 
Our $s^{NP}$ and $\bar{s}^{NP}$ distributions, as well as 
the difference between them, are similar in shape and magnitude
to those found in a recent analysis  
with a different approach based on two-body wave functions with
a number of parameters \cite{bro-ma}.
Another point deserving a comment concerns 
the probability of the strange meson-baryon fluctuation. 
As far as we know, direct experimental information is still not available 
and one then has to rely on phenomenological estimates that it should 
range between 4 and 10$\%$.

It has been recently shown that double polarization observables 
in $\phi$ 
meson photoproduction off protons \cite{phi} are very 
sensitive to the strange 
quark content of the proton itself, indicating an interesting source of 
experimental information for giving further support to the current discussion.

The present analysis can be easily extended to study non-perturbative 
contributions to other flavors. A pion cloud version of the model
should also predict a noticeable $\bar{u}-\bar{d}$ 
asymmetry in the sea of the nucleon, as announced by the recent 
data of the E866 Collaboration \cite{e866}. In this sense, 
it is worth to mention that the data analysis resulting from the above 
mentioned experiment favors an explanation in terms of the MCM rather than 
some alternative approaches. 

%%%%%%%%%%%%%%%%%%%%%%%%%%%

\section*{Acknowledgments} 

%%%%%%%%%%%%%%%%%%%%%%%%%%%

We acknowledge useful discussions with M. Malheiro. 
We also thank Centro Brasileiro de Pesquisas F\'{\i}sicas (CBPF) for the 
warm hospitality extended to us during this work. 
H.R.C. was partially supported by Centro Latino Americano de Fisica
(CLAF). J.M. and H.R.C. are supported by Funda\c{c}\~ao de Amparo 
\`a Pesquisa do Estado de Rio de Janeiro (FAPERJ).

%%%%%%%%%%%%%

%

\newpage

\section*{Figure Captions}
 
\begin{itemize}
\item
[Fig. 1:] Meson and baryon probability densities in the nucleon,
normalized to 4$\%$. 
Full line is for the strange meson distribution in the nucleon. Dashed 
line shows the strange baryon distribution $P_B(x)=P_M(1-x)$.

\item
[Fig. 2:] Upper: The non-perturbative strange distributions in the nucleon sea. 
Full lines are for $s^{NP}$ (thin) and $\bar{s}^{NP}$ 
(thick), as obtained by using eqs.~(10) and (11). 
Point lines are for the non-perturbative distributions coming from
the choice 
$s_B = u_N/2$ (thin) and $\bar{s}_M = (1-x)^{0.18}q^\pi$ (thick). 
\newline
Lower: The $s(x)-\bar{s}(x)$ asymmetry in the nucleon sea. 
The full line is the asymmetry obtained by using $s_B$ and $\bar{s}_M$ 
as given by eqs.~(10) and (11), whereas the point line
is the asymmetry coming from a $s_B = u_N/2$ distribution in the 
hyperon and a  $\bar{s}_M = (1-x)^{0.18}q^\pi$ in the kaon.
\end{itemize}

%\end{document}

\newpage
\begin{figure}[b] 
\psfig{figure=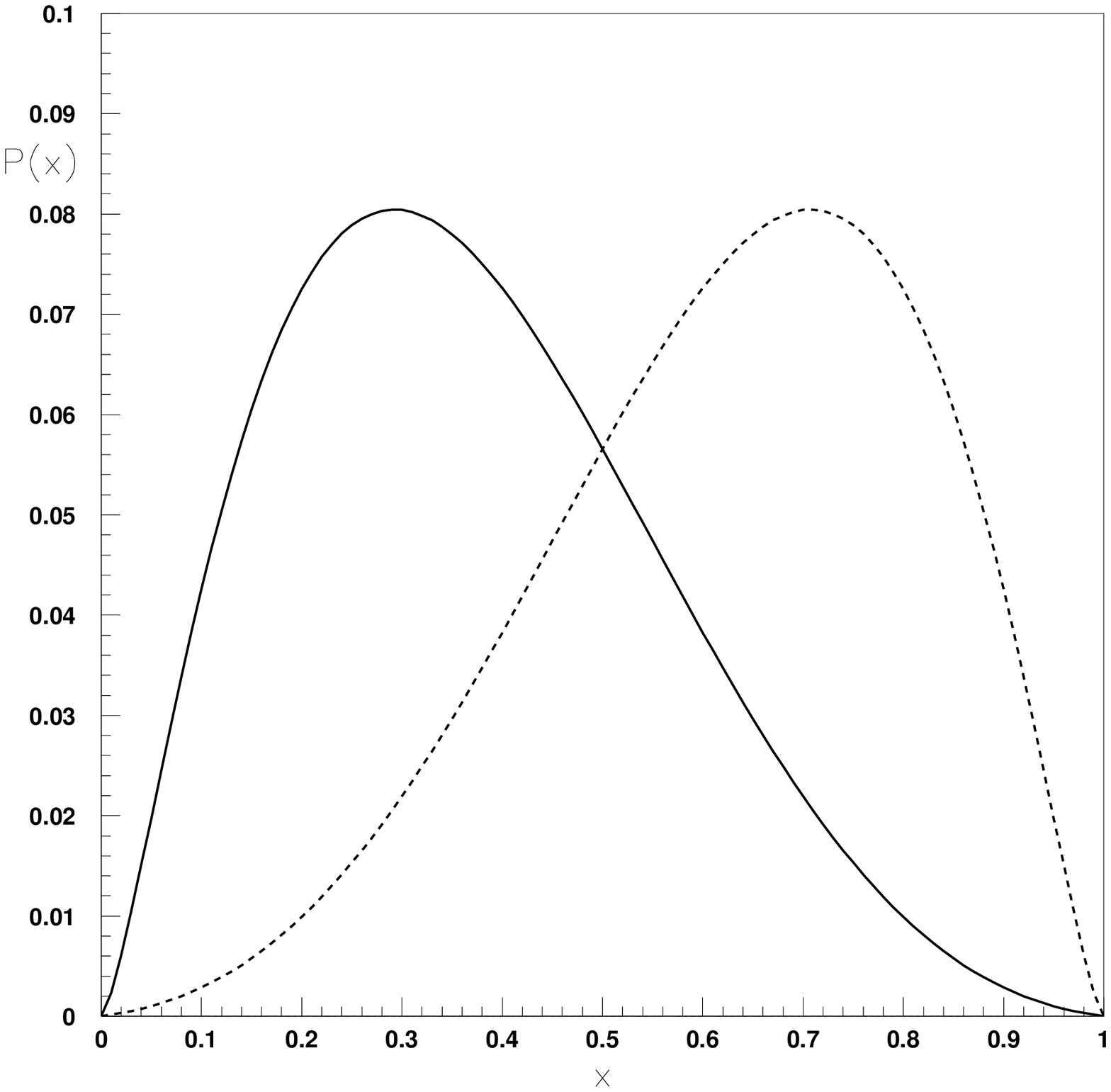,height=6.0in}
\caption{} 
\label{fig1}
\end{figure}
\begin{figure}[b] 
\psfig{figure=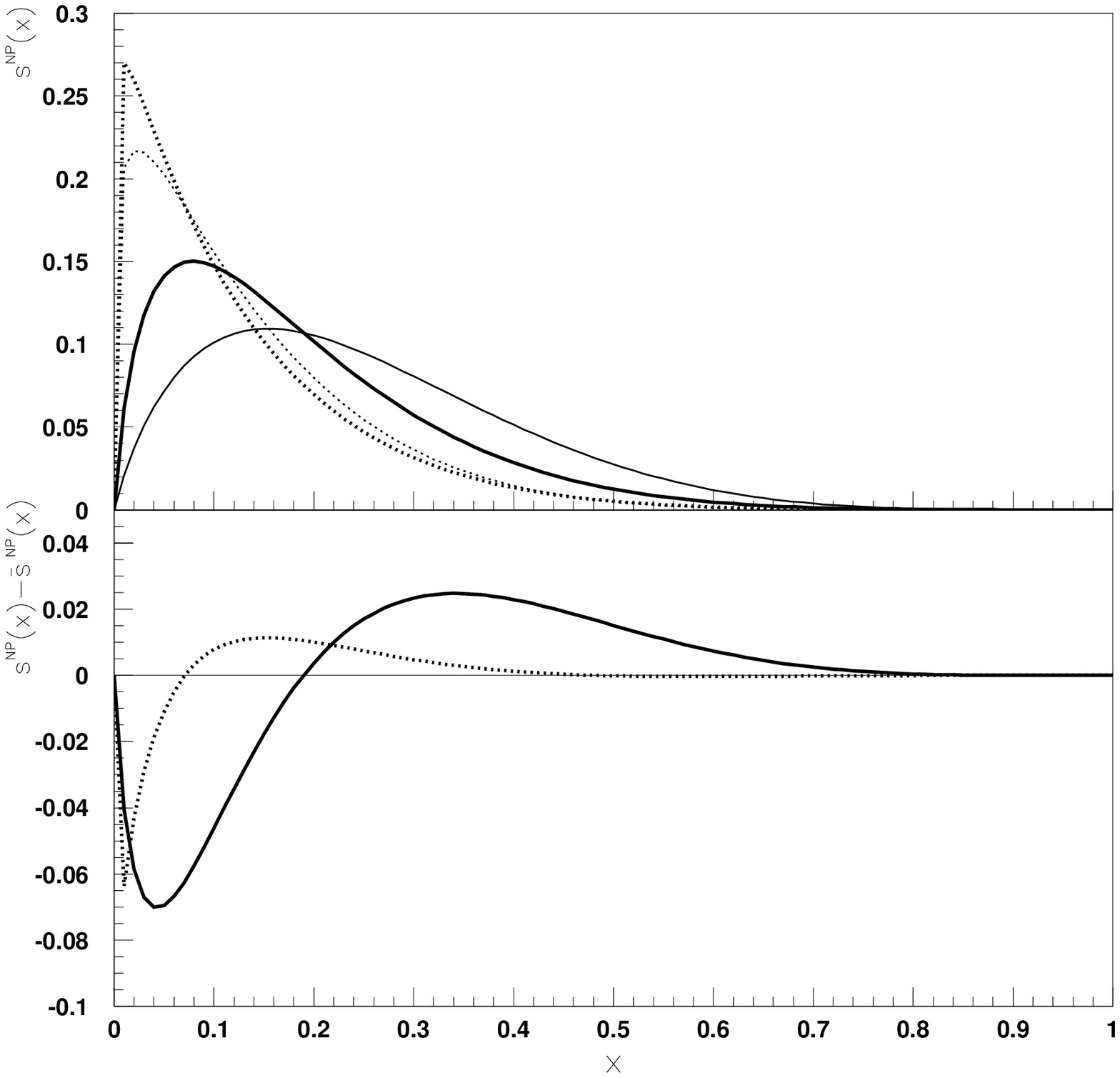,height=6.0in}
\caption{} 
\label{fig2} 
\end{figure}


\begin{thebibliography}{99}

\bibitem{mel-tho} W. Melnitchouk and A.W. Thomas, Phys. Lett. 
{\bf B414}, 134 (1997).

\bibitem{h1} C. Adloff {\it et al.} (H1 Colaboration), Z. Phys. {\bf C74}, 
191 (1997).

\bibitem{zeus} J. Breitweg {\it et al.} (ZEUS Collaboration), Z. Phys. 
{\bf C74}, 207 (1997).

\bibitem{ccfr} A.O. Bazarko {\it et al.} (CCFR Collaboration), 
Z. Phys. {\bf C65}, 189 (1995).

\bibitem{cteq} J. Botts {\it et al.} (CTEQ Collaboration),
 Phys. Lett. {\bf 304}, 159 (1993);
H. Lai {\it et al.} Phys. Rev. {\bf D51}, 4763 (1995).

\bibitem{signal} A. Signal and A.W. Thomas, Phys. Lett. {\bf B191}, 
205 (1987).

\bibitem{bro-ma} S.J. Brodsky and B.Q. Ma, Phys. Lett. {\bf B381}, 
317 (1996). B.Q. Ma and S.J. Brodsky, SLAC-PUB-7501, hep-ph/9707408.

\bibitem{e866} E.A. Hawker {\it et al.} (E866/NuSea Collaboration), 
Phys. Rev. Lett {\bf 80}, 3715 (1998); J.C. Peng {\it et al.} 
(E866/NuSea Collaboration), Phys. Rev. {\bf D58}, 092004 (1998).

\bibitem{burkardt} M. Burkardt and and B.J. Warr, Phys. Rev. {\bf D45}, 
958 (1992).

\bibitem{hwa} R.C. Hwa, Phys. Rev. {\bf D22}, 759 (1980); 
{\it ibid.} 1593.

\bibitem{alta-par} G. Altarelli and G. Parisi, Nuc. Phys. {\bf B126}, 
298 (1977).

%\bibitem{grv} M. Gl\"uck, E. Reya and A. Vogt, Z. Phys. {\bf C53}, 127 
%(1992).
%
\bibitem{das-hwa} K.P. Das and R.C. Hwa, Phys. Lett {\bf B68}, 459 (1977).

\bibitem{malheiro} W. Melnitchouk and M. Malheiro, Phys. Rev. {\bf C55}, 
431 (1997).

\bibitem{shigetani} T. Shigetani, K. Suzuki and H. Toki, Phys. Lett. 
{\bf B308}, 383 (1993). 

\bibitem{kaon} J.T. Londergan, G.Q. Liu and A.W. Thomas, 
Phys. Lett. {\bf B380}, 393 (1996).

\bibitem{suecos} A. Edin and G. Ingelman, Phys. Lett. {\bf B432}, 
402 (1998).

\bibitem{phi} A.I. Titov, Y. Oh, S.N. Yang and T. Morii, Phys. Rev. 
{\bf C58}, 2429 (1998).

\end{thebibliography}
\end{document}